\documentclass[twocolumn]{aastex63}

\usepackage{graphicx}
\usepackage{color}
\usepackage{amsmath}  
\usepackage{verbatim}     
\usepackage{hyperref}
\usepackage[noabbrev,capitalise]{cleveref}  
\usepackage{natbib}
\usepackage[mathscr]{euscript} 
\usepackage{xspace}

\makeatletter
\usepackage{etoolbox}
\patchcmd\H@refstepcounter{\protected@edef}{\protected@xdef}{}{}
\makeatother

\newcommand{\sn}{SN~Ia\xspace}
\newcommand{\sne}{SNe~Ia\xspace}
\newcommand{\snemo}{\textsc{SNEMO}\xspace}
\newcommand{\snemotwo}{\textsc{SNEMO2}\xspace}
\newcommand{\snemofive}{\textsc{SNEMO5}\xspace}
\newcommand{\snemosix}{\textsc{SNEMO6}\xspace}
\newcommand{\snemoseven}{\textsc{SNEMO7}\xspace}
\newcommand{\snemofifteen}{\textsc{SNEMO15}\xspace}
\newcommand{\salt}{\textsc{SALT2}\xspace}

\newcommand{\salttwofour}{\textsc{SALT2.4}\xspace}

\newcommand{\lcdm}{$\Lambda$CDM\xspace}     
\newcommand{\h}{\ensuremath{\text{H}_0}\xspace}

\newcommand{\un}[1]{~\text{#1}\xspace}  

\newcommand{\unity}{UNITY1.2\xspace}

\newcommand{\qone}{When applying the model to current light-curve-only data, are the standardization coefficients consistent with those derived from the  \added{spectrophotometric time-series} training data set?\xspace}
\newcommand{\qoneshort}{Are the standardization coefficients consistent between data sets?\xspace}
\newcommand{\qtwo}{How many standardization coefficients are distinguishable from zero?\xspace}
 
\newcommand{\qthree}{What are the correlations between the coefficients?\xspace}

\newcommand{\qfour}{Given current data sets, does \snemoseven reduce the need for unexplained intrinsic scatter in \sn (\ensuremath{\sigma_{\rm unexplained}}) in the Hubble-Lema\^itre diagram?\xspace}

\newcommand{\qfive}{Does \snemoseven reduce the correlations with host-galaxy properties, such as the one with stellar mass (\ensuremath{\gamma})?\xspace}

\newcommand{\qfourfive}{Does \snemoseven reduce unexplained and systematic variations in standardization?}

\newcommand{\nsneTotal}{914\xspace}

\newcommand{\nsneSaltSnemoTwo}{867\xspace}

\newcommand{\nsne}{240\xspace} 
\newcommand{\nsneOnePercentOneSigma}{90\xspace}
\newcommand{\nsneOnePercent}{194\xspace} 

\newcommand{\nsneTwoPercent}{126\xspace} 

\newcommand{\sigmaiNorm}{2\xspace}
\newcommand{\ciMax}{5\xspace}

\newcommand{\saltdeltaSnemoTwo}{\ensuremath{-0.043 \pm 0.010}\xspace}

\newcommand{\intrinicScatterSnemoTwoTwoError}{\ensuremath{0.102 \pm 0.008~\text{mag}}\xspace} 

\newcommand{\saltdelta}{\ensuremath{-0.01 \pm 0.02}\xspace} 
\newcommand{\intrinicScatterSaltZeroTwo}{\ensuremath{0.135 \pm 0.009~\text{mag}}\xspace} 
\newcommand{\intrinicScatterSaltTwoTwo}{\ensuremath{0.141 \pm 0.015~\text{mag}}\xspace} 
\newcommand{\intrinicScatterSevenZeroTwo}{\ensuremath{0.125 \pm 0.011~\text{mag}}\xspace} 
\newcommand{\intrinicScatterSevenTwoTwo}{\ensuremath{0.10 \pm 0.03~\text{mag}}\xspace} 

\newcommand{\percentOutliersMin}{1.5\%\xspace}
\newcommand{\percentOutliersTypicalMax}{3.2\%\xspace}

\newcommand{\saltrms}{\ensuremath{0.148~\text{mag}}\xspace}
\newcommand{\snemorms}{\ensuremath{0.141~\text{mag}}\xspace}

\begin{document}

\title{Initial Evaluation of SNEMO2 and SNEMO7 Standardization Derived From Current Light Curves of Type Ia Supernovae}
\shorttitle{Initial Evaluation of SNEMO Light Curve Standardization}

\author[0000-0002-1873-8973]{B. M. Rose}
\affiliation{Space Telescope Science Institute, 3700 San Martin Drive
Baltimore, MD 21218}
\author[0000-0003-1861-0870]{S. Dixon}
\affiliation{E.O. Lawrence Berkeley National Laboratory, 1 Cyclotron Rd., Berkeley, CA, 94720}
\affiliation{Department of Physics, University of California Berkeley, 366 LeConte Hall MC 7300, Berkeley, CA, 94720-7300}
\author[0000-0001-5402-4647]{D. Rubin}
\newcommand{\uhawaii}{\affiliation{Department of Physics and Astronomy, University of Hawai`i at M{\=a}noa, Honolulu, Hawai`i 96822}}
\uhawaii
\affiliation{Space Telescope Science Institute, 3700 San Martin Drive
Baltimore, MD 21218}
\affiliation{E.O. Lawrence Berkeley National Laboratory, 1 Cyclotron Rd., Berkeley, CA, 94720}
\author[0000-0002-0476-4206]{R. Hounsell}
\affiliation{University of Pennsylvania Department of Physics \& Astronomy, 209 South 33rd Street, Philadelphia, PA 19104}
\author[0000-0002-4094-2102]{C. Saunders}
\affiliation{Sorbonne Universit\'e, Universit\'e Paris Diderot, CNRS/IN2P3, Laboratoire de Physique Nucl\'eaire et de Hautes \'Energies, 4 Place Jussieu, Paris, France}

\collaboration{5}{ }

\author[0000-0003-2823-360X]{S. Deustua}
\affiliation{Space Telescope Science Institute, 3700 San Martin Drive
Baltimore, MD 21218}
\author[0000-0002-6652-9279]{A. Fruchter}
\affiliation{Space Telescope Science Institute, 3700 San Martin Drive
Baltimore, MD 21218}
\author[0000-0002-1296-6887]{L. Galbany}
\affiliation{Physics and Astronomy Department, University of Pittsburgh, 304 Allen Hall, 3941 O'Hara St, Pittsburgh PA 15260}
\affiliation{Departamento de F\'isica Te\'orica y del Cosmos, Universidad de Granada, E-18071 Granada, Spain}
\author{S. Perlmutter}
\affiliation{E.O. Lawrence Berkeley National Laboratory, 1 Cyclotron Rd., Berkeley, CA, 94720}
\affiliation{Department of Physics, University of California Berkeley, 366 LeConte Hall MC 7300, Berkeley, CA, 94720-7300}
\author[0000-0003-2764-7093]{M. Sako}
\affiliation{University of Pennsylvania Department of Physics \& Astronomy, 209 South 33rd Street, Philadelphia, PA 19104}

\correspondingauthor{B. M. Rose}
\email{brose@stsci.edu}
\shortauthors{Rose, Dixon, Rubin, et al.}

\received{September 25, 2019}
\revised{December 20, 2019}
\submitjournal{The Astrophysical Journal}

\begin{abstract}
To determine if the SuperNova Empirical Model (SNEMO) can improve Type Ia supernova (SN Ia) standardization \replaced{with the}{of} several currently available photometric data sets, we perform an initial test, comparing results with the much-used SALT2 approach. We fit the SNEMO light-curve parameters and pass them to the Bayesian hierarchical model UNITY1.2 to estimate the Tripp-like standardization coefficients, including a host mass term as a proxy for \added{redshift dependent} astrophysical systematics\deleted{ that could affect cosmology measurements}. We find that, among the existing large data sets, only the Carnegie Supernova Project data set consistently provides the signal-to-noise and time sampling \replaced{that can}{necessary to} constrain the additional five parameters that SNEMO7 incorporates beyond SALT2. This is an important consideration for future SN Ia surveys like LSST and WFIRST. Although the SNEMO7 parameters are poorly constrained by most of the other available data sets of light curves, we find that the SNEMO2 parameters are just as well-constrained as the SALT2 parameters. In addition, SNEMO2 \added{and SALT2} \replaced{has}{have} comparable \added{unexplained} intrinsic scatter \deleted{on the Hubble-Lema\^itre diagram} when fitting the same data\deleted{as SALT2}. \added{When looking at the total scatter, SNEMO7 reduces the Hubble-Lema\^itre diagram RMS from 0.148~mag to 0.141~mag.} It is not then, the SNEMO methodology, but the interplay of data quality and the increased number of degrees of freedom that is behind these \replaced{poorer}{reduced} constraints. With this in mind, we recommend further investigation into \added{the data required to use SNEMO7 and} the possibility of fitting the poorer photometry data with intermediate SNEMO-like models with three to six \deleted{principal}components.
\end{abstract}


\section{Introduction}\label{intro}

\explain{We are now tracking changes via the AASTeX tools, \url{https://journals.aas.org/aastexguide/\#revised_color}.} 

Observations show that measurable properties of Type Ia supernovae (\sne) are correlated with their peak brightnesses, making \sne standardizable candles \citep{Phillips1993,Hamuy1996d,Riess1996,Perlmutter1997}. Once their peak brightnesses are standardized, they can be used as distance indicators and aid in our understanding of the expansion history of universe. The precision with which cosmological parameters are constrained depends, in part, on how well \sn standardization reduces the dispersion in peak brightnesses. Beginning in the 1990s, standardization techniques were developed that reduced the dispersion to 0.15 mag, resulting in the  discovery of the accelerating expansion of the universe \citep{Riess1998,Perlmutter1999}. Since then, improving techniques for \sn standardization has been a continuous topic of research. \citep[e.g.][]{Phillips1999,Guy2005,Jha2007,Guy2007,Burns2011}. Further improvements may be needed in order to remove possible percent level systematics \citep[e.g.][]{Foley2011,Kim2013,Fakhouri2015,Pierel2018,Burns2018,Hayden2019} that could affect \added{systematic uncertainty limited} measurement\added{s} of Dark Energy by future missions like LSST \citep{LSSTScienceCollaboration2009} and WFIRST \citep{Spergel2015, Hounsell2018}.

\replaced{\sne are transients that are typically observed photometrically with a cadence of several days, resulting in multiple broadband light curves.}{\sne are typically observed photometrically in a few broadband optical filters with an observation every few days.} \replaced{These l}{The resulting l}ight curves are then fit to one of several empirically-based models in order to extract \sn parameters that quantify properties like light-curve shape and \deleted{intrinsic} color. These parameters are then used to standardize the absolute luminosity of the supernovae \replaced{ and these standardized luminosities are used to measure distances}{, measure distances,} and eventually constrain cosmological parameters. The exact interpretation of these parameters differs for each light-curve fitting method.

Light-curve fitters like \citet{Hamuy1996d}, \citet{Riess1996}, \citet{Phillips1999}, and \citet{Jha2007} use a single light-curve shape parameter and \replaced{deal with extinction from dust and intrinsic}{separate the sources of} \sn color variation\deleted{s} by assuming a fixed Milky Way\added{-like} extinction curve to describe the \added{variation due to} dust and attributing the remaining color variation to intrinsic color differences in the \sne.
\citet{Tripp1998} and \citet{Guy2007} also use \replaced{one}{a single} light-curve shape parameter, but do not separate the sources of color variation\deleted{s}.
The popular \salt \added{model} \citep{Guy2007,Guy2010,Betoule2014,Mosher2014} uses a linear model of the \sn spectral energy distribution sequence fit from light curves and spectra. The \replaced{light curves are}{model is} parameterized by finding the coefficients \deleted{of the model} that produce\deleted{s} synthetic photometry most similar to the observed photometry. One parameter, $x_1$, captures the broader-brighter (or Phillips) relationship identified in \citet{Phillips1993} and \citet{Pskovskii1977}. For normal \sne, \added{the distribution of} $x_1$ roughly follows a standard normal distribution. The second parameter, $c$, accounts for color variability both from dust and intrinsic diversity. For typical \sne, $c$ is \added{also roughly normally  distributed, but with a narrower spread; most \sne have a $c$ value} within a few tenths of a magnitude of zero. 

The standardization method commonly referred to as Tripp standardization \citep{Tripp1998}, combines these light-curve shape and color parameters linearly to \replaced{standardize \sne}{estimate the distance modulus, $\mu$}. \replaced{(typically standardizing the rest-frame $B$-band magnitude at maximum light, $m_B$)}{This is typically done for the rest-frame $B$-band magnitude ($m_B$).} Using the parameters from the \salt \sn light-curve model, the Tripp standardization equation is:
\begin{equation}\label{eqn:salt}
\mu = m_B - (M_B - \alpha\; x_1 + \beta\; c)
\end{equation}
where $\mu$, $m_B$, $M_B$ are the distance modulus, apparent magnitude, and absolute magnitude respectively. The $\alpha$ and $\beta$ parameters are the linear standardization coefficients corresponding to the \sn light-curve shape ($x_1$) and intrinsic color ($c$)\deleted{, respectively}. The parameters $m_B$, $x_1$, and $c$ are fit for each individual \sn, while $M_B$, $\alpha$, and $\beta$ are global parameters that are fit simultaneously, along with the cosmological parameters of interest, using the full data set.

There is evidence suggesting that \sne show considerably more spectral diversity than the \added{current} shape and color parameters capture \citep{Branch2006,Kim2013,Fakhouri2015,Hayden2019,Rubin2019}. This diversity may present itself as uncorrected systematic shifts in the peak luminosity of \sn. An example of such an \replaced{uncorrected}{unaccounted for} systematic is seen in the host galaxy mass step \citep{Kelly2010,Sullivan2010,Lampeitl2010}. The mass step is \replaced{the}{a} shift in average peak luminosity of $\sim 0.06 \un{mag}$\replaced{, when comparing}{  between} \sn from low stellar mass host galaxies ($\lesssim 10^{10}~ \text{M}_{*}/\text{M}_{\odot}$) to high mass hosts ($\gtrsim 10^{10}~ \text{M}_{*}/\text{M}_{\odot}$). This result has been seen in multiple samples with a $>5\sigma$ significance \citep{Childress2013b,Uddin2017a,Moreno-Raya2018}.

\subsection{SNEMO}

In order to address the\replaced{se issues}{ issue of unmodeled spectral diversity}, \citet{Saunders2018} presented the Super-Nova Empirical MOdels (\snemo), which applies expectation maximization factor analysis (EMFA, a dimensionality reduction algorithm similar to principal component analysis) to optical spectrophotometric time series \added{data} obtained by the Nearby Supernova Factory (SNfactory, \citealt{Aldering2002}). 
EMFA reduces the dimensionality of the training data set to a predefined number of eigenvectors. In the case of \snemo, these eigenvectors are time series of spectra \citep[Equations 7 \& 10]{Saunders2018}. Combined, these eigenvectors represent a linear basis from which one can reconstruct any optical \sn spectral time series\deleted{ model}. This method of defining the eigenvectors is similar to the method used to define \salt's $x_1$ \citep[Section 5]{Guy2007}, however \added{the EMFA algorithm used to obtain the} \snemo \added{components} handles missing and noisy data in a different manner and does not use any photometric \added{training} data.
\added{Additionally,} \snemo does not fit a variable color law and instead assumes a \citet{Fitzpatrick2007} reddening law \citep[Section 3.2]{Saunders2018}.
Like \salt, each of the best-fit model coefficients (or eigenvector projections) describe a certain light-curve shape and can be combined to standardize supernova magnitudes. Unlike some light-curve shape parameters (e.g. $\Delta m_{15}$), these EMFA eigenvectors are pure mathematical constructs and do not necessarily connect to anything physical or intuitive. 

\snemo is a family of models\replaced{, each using a different number of eigenvectors}{ trained on the same data}.
\deleted{Using the same training data, models were created with many numbers of eigenvectors.}
\citet{Saunders2018} released three variants\footnote{\url{https://snfactory.lbl.gov/snemo/index.html}}: \snemotwo, \snemoseven, and \snemofifteen\footnote{Unlike principal component analysis, when using the same data to generate models with differing numbers of eigenvectors, EMFA does not guarantee that the first few eigenvectors are the same. That means \snemoseven is not just the first seven eigenvectors of \snemofifteen. However, in practice the first three or four eigenvectors of these two models are nearly identical.}. \snemotwo is named for its two spectral-temporal eigenvectors; \snemoseven and \snemofifteen have seven and fifteen eigenvectors respectively. In addition, each \snemo model has a color correction curve that is identical to the \citet{Fitzpatrick2007} reddening law. The ``zeroth" eigenvector, describing the mean spectral-temporal evolution, is related to $m_B$ in \cref{eqn:salt}, and its corresponding coefficient used only as an overall scaling factor. The other spectral-temporal and color parameters are combined linearly to standardize \sne.

\snemotwo, which consists of a mean vector, one spectral-temporal component of variation, and a color law, is directly analogous to \salt, differing only in the training data and methodology \added{used to obtain the model components}. \snemotwo allows \added{for} a direct comparison between the \snemo and \salt training methodologies without introducing any more degrees of freedom \added{to the model}. The other \snemo models introduce more parameters to allow the model to capture more of the spectral variation. In the initial release of \snemo, \citet{Saunders2018} showed that these extra parameters do improve the quality of the model in fitting the diversity of \sn behavior.

\replaced{Considering both}{Using} the SNfactory training and a separate SNfactory validation set, \snemofifteen was found to be the model able to capture the most spectral diversity while avoiding overfitting. \snemoseven was considered to be a model that well-sampled multi-band light curves should be able to constrain, while capturing more \sn variation than \snemotwo (or \salt). In addition, \snemoseven was determined to be the point of diminishing returns when using Tripp-like linear standardization. It is worth noting that there is evidence for \added{the further consideration of} non-linear \added{spectral} behavior \added{(e.g. ejecta velocities)} that may require the \replaced{better}{more descriptive spectral} fits \replaced{of}{obtained with} \snemofifteen.

\begin{figure*}
    \centering
    \includegraphics[width=0.9\textwidth]{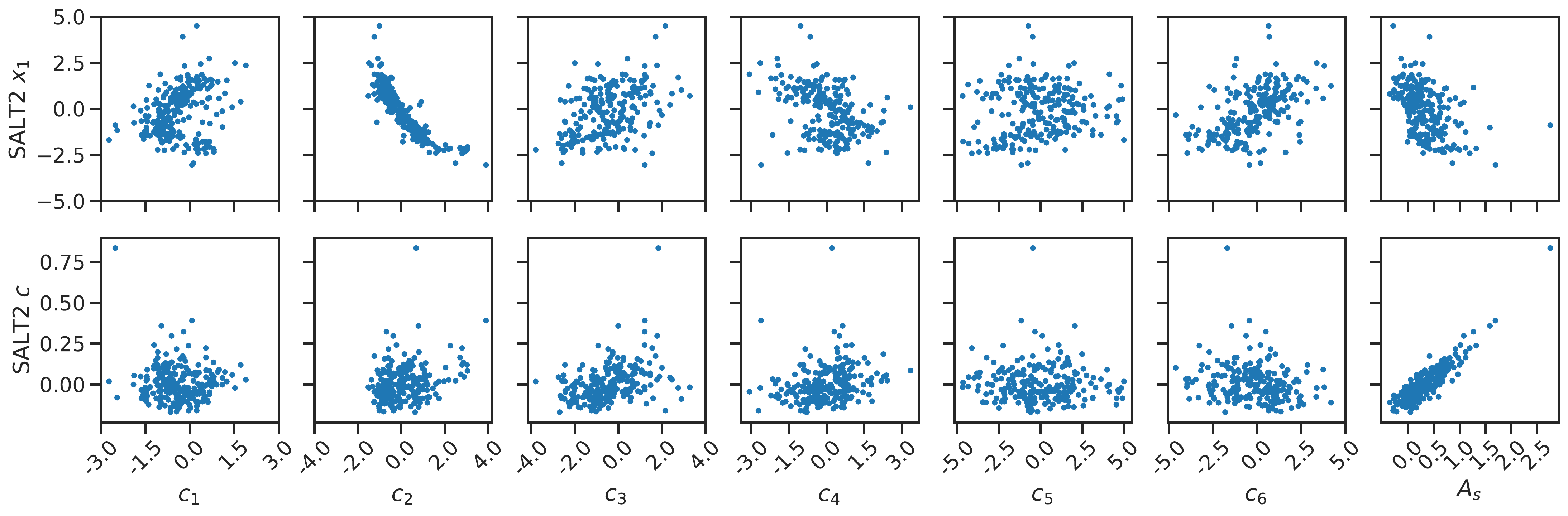}
    \caption{\added{Correlations between parameters of \salt and \snemoseven from our nominal photometric data set: no error model, $\sigma_i \leq \sigmaiNorm$, and without outliers (N=229).  This is similar to the \snemotwo correlations from spectro-photometric data shown in Figure 13 of \citet{Saunders2018}.}
    }
    \label{fig:salt_vs_snemo}
\end{figure*}

\added{For a further understanding of the similarities and difference between \salt and \snemo, we plot the correlations between the model parameters in \cref{fig:salt_vs_snemo}. Figure 13 of \citet{Saunders2018} shows this same plot but for \snemotwo parameters measured from spectro-photometric data. This work focuses on \snemoseven parameters derived from photometric data, so \cref{fig:salt_vs_snemo} uses our nominal \snemoseven data set. Details on the data and model are described in Section \ref{sec:DataAndUNITY}.}

In this work, we perform the initial test of how well \snemoseven standardizes \sne using only publicly available photometric light-curve data. This goes beyond the spectrophotometric time series data set used in \replaced{its}{the} development and initial testing \added{of the \snemo models}. We include a host stellar mass term as a proxy for any uncorrected \sn astrophysical systematics. It is these possible unknown systematics that stand as the largest threat to precision Dark Energy measurements.
Host stellar mass \deleted{is used because since the detection of the effect by it}has become a standard proxy \added{since \citet{Kelly2010}}. However more recent research by \citet{Gupta2011}, \citet{Hayden2013}, \citet{Rigault2013}, \citet{Childress2013b}, \citet{Childress2014}, \citet{Moreno-Raya2018}, \citet{Rigault2018}, \citet{Rose2019}, and others show that alternative astrophysical measurements may better match the true physical mechanism.

We use the following criteria to evaluate \snemoseven's ability to standardize current \added{photometric \sne} data sets\deleted{ of light curve only \sne}:
\begin{enumerate}
\item \qone 
\item \qtwo
\item \qthree Strong correlations imply that a projection needs to be fit even if the standardization coefficient is consistent with zero.
\item \qfour
\item \qfive A reduction of these correlations would imply a reduced systematic floor for \sn standardization.
\end{enumerate}
These tests do not attempt to validate or characterize \snemo's ability to fit light curves, but rather focus on questions concerning \added{population-level effects that have the potential to impact} cosmological measurements. Characterizing light curve fits will be done thoroughly in a forthcoming paper (Saunders et al. 2020). We will also not investigate the limits of Tripp-like standardization equations or methods. Finally, we are here only asking how \snemo fares on these 5 criteria when given the current quality of \sn data sets, not how it performs when given the data quality expected from\deleted{, e.g.,} LSST \replaced{and}{or} WFIRST. These are all important research topics and should be discussed independently.

In \cref{sec:DataAndUNITY}, we discuss the photometric data and the method we used to test \snemoseven, and
in \cref{sec:Results}, we discuss our findings and present answers the five questions above. \added{In Section \ref{sec:impacts}, we discuss how these results impact \sn cosmology in the systematics dominated era of LSST and WFIRST.}

\section{The Data and UNITY}\label{sec:DataAndUNITY}

This work uses high-redshift ($z \gtrsim 0.5$) Hubble Space Telescope (HST) data from \citet{Riess2007},
mid-redshift ($0.1 \lesssim z \lesssim 0.4$) data from the rolling supernovae surveys of the Sloan Digital Sky Survey (SDSS, \citealt{Sako2014}) and the ($0.1 \lesssim z \lesssim 1.0$) Supernova Legacy Survey (SNLS, \citealt{Betoule2014}),
and nearby ($z<0.1$) \sne observed with targeted followup from the Foundation survey \citep{Foley2018}, the Carnegie Supernova Project (CSP) third data release \citep{Krisciunas2017}, and the Center for Astrophysics Fred Lawrence Whipple Observatory Supernovae data releases \citep[CfA,][]{Riess1999, Jha2006, Hicken2009, Hicken2012}. Of these many data sets, CSP followed the \sne at a faster cadence than most and obtained observations with higher-than-typical signal-to-noise.
We use only objects with available host galaxy stellar mass measurements (Gupta et al. 2020). The number of \sne from each survey are in \added{the first two rows of} \cref{tab:LcFitData}. The total size of the sample with host galaxy stellar mass measurements is \nsneTotal.

\begin{deluxetable*}{l|cccccc|c}[t]
\tablecolumns{7}
\tabletypesize{\footnotesize}  
\tablewidth{3in}
\tablecaption{Number of \sn passing quality cuts from various SNEMO models.\label{tab:LcFitData}} 
\tablehead{
    \colhead{} &
    \colhead{CSP} & \colhead{Foundation} & \colhead{CfA} &
    \colhead{SDSS} & \colhead{SNLS} & \colhead{HST} &
    \colhead{Total}
    }
\startdata
Total SNe & 134 & 223 & 97 & 371 & 239 & 9 & 1073\\
Host mass avail. & 99 & 99 & 97 & 371 & 239 & 9 & 914\\
\hline
& & \multicolumn{4}{c}{\hspace{1em}\snemotwo} & & \\
\textbf{No Error Model} & & & & & & \\
\hspace{1em}$\sigma_i \leq 2$ & 96 &  99 & 95 & 355 & 234 & 6 & 885\\
\textbf{1\% Error Model} & & & & & & \\
\hspace{1em}$\sigma_i \leq 2$ & 96 &  98 & 96 & 355 & 234 & 7 & 886\\
\textbf{2\% Error Model} & & & & & & \\
\hspace{1em}$\sigma_i \leq 2$ & 97 & 98 & 94 & 352 & 234 & 6 & 881\\
\hline
& & \multicolumn{4}{c}{\hspace{1em}\snemoseven} & & \\
\textbf{No Error Model} & & & & & & \\
\hspace{1em}$\sigma_i \leq 1$ & 80 &  36 & 16 & 12 & 13 & 0 & 157\\
\hspace{1em}$\sigma_i \leq 2$ & 83 &  62 & 45 & 24 & 26 & 0 & 240\\
\textbf{1\% Error Model} & & & & & & \\
\hspace{1em}$\sigma_i \leq 1$ & 66 &  9 & 4 & 11 & 0 & 0 & 90\\
\hspace{1em}$\sigma_i \leq 2$ & 75 &  52 & 28 & 21 & 18 & 0 & 194\\
\textbf{2\% Error Model} & & & & & & \\
\hspace{1em}$\sigma_i \leq 1$ & 36 &  0 & 1 & 2 & 0 & 0 & 39\\
\hspace{1em}$\sigma_i \leq 2$ & 73 &  22 & 9 & 15 & 7 & 0 & 126\\
\enddata
\tablecomments{$\sigma_i$ is the uncertainty on each fit eigenvector. When $\sigma_i=1$, the uncertainty is approximately the $1\sigma$ dispersion in the population. The data from the CfA, SDSS, SNLS, and HST surveys were obtained via the JLA compilation \citep{Betoule2014}.}
\end{deluxetable*}

\subsection{\snemoseven Light-Curve Fits}\label{sec:snemo7}

The \added{three released} \snemo models are available in the \verb|sncosmo| python package\footnote{\doi{10.5281/zenodo.592747}} (version 1.7). 
We use the \verb|mcmc_lc| function in that package to find the posterior distribution of the best-fit model coefficients (i.e. the eigenvector projections, $c_i$) \added{for each \sn} and estimate their uncertainties ($\sigma_i$) from these posteriors. This function uses MCMC to sample from
\begin{equation} \label{eq:chisquared}
\chi^2 = {(\mathbf{f}_\mathrm{obs} - \mathbf{f}_\mathrm{mod})^\top}(\Sigma_\mathrm{obs}+\Sigma_\mathrm{mod})^{-1}(\mathbf{f}_\mathrm{obs} - \mathbf{f}_\mathrm{mod}) \;.
\end{equation}
This is a function of the model coefficients $(z, t_0, c_i, A_s)$, where $\mathbf{f}_{\mathrm{obs}}(b, p)$ is the flux observed in bandpass $b$ at phase $p$ and $\mathbf{f}_{\mathrm{mod}}(b, p; z, t_0, c_i, A_s)$ is the flux predicted in bandpass $b$ at phase $p$ obtained by performing synthetic photometry on the spectral time series model with the given model coefficients. A diagonal covariance matrix \replaced{made up of}{whose entries represent} the observational uncertainty \added{in each bandpass and phase observed} ($\Sigma_{\mathrm{obs}}$) is added to the model covariance ($\Sigma_{\mathrm{mod}}$) \added{to obtain the full covariance matrix used in the light curve fits}.
The population dispersion of a given component ($c_i$) is normalized to approximately 1, meaning $\sim$ 1,000 normal \sne should have a $c_i$ range of $\sim$ $-3$ to $3$. Further details on the interpretation of \snemo parameters can be found in \citet{Saunders2018}. 
In addition to the \snemo coefficients, the time of maximum brightness is fit along with the model coefficients with wide, uniform priors ($(-50, 50)$ for each of the model coefficients, and $(\min(t_{\mathrm{obs}})-20, \max(t_{\mathrm{obs}}))$ for the time-of-max). When running the inference, we let the redshift in \snemo vary within the uncertainty of the measurement ($\sim$ 0.0001). We also correct for Milky Way dust reddening using the \cite{Schlafly2011} maps.

In a \salt-like analysis, initial light-curve quality cuts based on phase sampling and signal-to-noise are usually applied. 
We do not yet have a similar heuristic for which light curves are high enough quality to \added{be} fit with \snemoseven.
Instead, we filter the \sne on the \snemoseven parameters and uncertainties directly, rather than any other measured properties of the light curves.
We define an object to be well fit by \snemoseven when its eigenvector coefficient values are less than a threshold ($|c_i| < \ciMax$) and the uncertainties on those coefficients are also smaller than another threshold ($\sigma_i \leq \sigmaiNorm$).
The cut on $|c_i| < \ciMax$ is intended to remove large outliers, and the cut on $\sigma_i > \sigmaiNorm$ removes \sn that have an uncertainty in the best-fit coefficients larger than twice the $1\sigma$ population dispersion.
A high $\sigma_i$, e.g. $> 2$, represents data that can be fit by a wide range of models, effectively putting no constraint on the true values of the model parameters.
We investigate the effect of different $\sigma_i$ cutoff values on our results in Appendix~\ref{sec:SNRCut}.
These quality cuts remove unconstrained fits without excessively restricting our sample size. 
This results in a \added{nominal} data set of $N = \nsne$.

\snemoseven does not yet have an uncertainty model (the $\Sigma_{\mathrm{mod}}$ in Equation~\ref{eq:chisquared}). A formal uncertainty model describes the regions in parameter space where \sne are more diverse than the model and reduces the impact these regions have on fitting data. \replaced{A formal}{This} uncertainty model is under development (Saunders et al. 2020), but \added{in this work} we \deleted{do} need to look at the effect of treating the model as imperfect.
For \salt, the uncertainty model is partially determined by the statistical uncertainty from their training data set \citep{Guy2007}; the model is more certain in areas that had more training data. For \snemo, the training data was selected to all have the same rest-frame wavelength coverage. As such, this part of the uncertainty model should be smooth. The other part of the SALT2 uncertainty model describes correlated residuals around the model. We expect this component \replaced{is}{to be} reduced for \snemoseven, as it describes more of the intrinsic SN behavior.
Using these assumptions, we investigate the effects of an imperfect model using a simplified uncertainty model. This naive uncertainty model consists of a diagonal covariance matrix with entries given by 1\% or 2\% of the peak flux value in each band. The formal uncertainty model in development, has more variation in phase than our naive model, but its scale is with in this range.

The addition of these uncertainties degrades the coefficient measurement precision \added{(i,e. $\sigma_i$)}, therefore reducing the number of \sn passing quality cuts.
With a 1\% naive uncertainty model, the number of \sn passing our quality cuts are $N=\nsneOnePercent$, and dropping to $N=\nsneTwoPercent$ with the 2\% uncertainty model. Ultimately, \replaced{the final data sets}{the data sets that survive these cuts} are dominated by CSP \sne. \added{Table \ref{tab:LcFitData} shows how varying the light-curve fit quality cuts and uncertainty model affects the total number of \sn in our sample.}

Several factors contribute to the poor constraints on the model parameters. A large factor is wavelength coverage. The \snemo model is defined from 3300--8600~\AA, and any observations in bands with rest-frame wavelengths outside of this range are not used to constrain the model parameters. As an example, we find that all of the SNLS objects that pass our cuts are at redshifts below $\sim0.7$, which is where the effective wavelength of the $r$-band falls below the lower bound of the \snemo wavelength range.
\replaced{Table \ref{tab:LcFitData} shows how varying the light-curve fit quality cuts and uncertainty model affects the total number of \sn in our sample}{The signal-to-noise ratio of the observations or the temporal sampling of the light curves can also have an impact on our ability to constrain the model parameters in the light curve fits. A full study of these effects is left to future work.}

\subsection{\unity}

We used the Unified Nonlinear Inference for Type Ia cosmologY (UNITY) framework to estimate the \deleted{coefficients our} standardization equation\added{, Equation \ref{eqn:unity} below}. UNITY, a Bayesian hierarchical model implemented in Stan \citep{Carpenter2017} using \verb|pystan| (\doi{10.5281/zenodo.598257}), was developed by \citet{Rubin2015} and further refined by \citet{Hayden2019}.  A more recent version (\unity) now includes the capability of modeling Tripp-like standardization equations with an arbitrary number of standardization parameters\deleted{, going beyond the standard two}.\footnote{These latest updates can be found at \url{https://github.com/rubind/host_unity}. The computational analysis procedures for this work are documented in \deleted{the} \texttt{rdr2019/makefile}.} 
Because our focus is on standardization and not cosmology directly, we assume a flat, \lcdm cosmology with $\Omega_M = 0.3$.

\added{All SED models considered (\salt, \snemotwo, and \snemoseven) are unable to achieve a dispersion in distance modulus that is consistent with measurement uncertainties and linear standardization. We model the remaining ``unexplained'' dispersion with a model parameter: unexplained intrinsic scatter, $\sigma_{\mathrm{unexplained}}$. We assume $\sigma_{\mathrm{unexplained}}$ describes the width of a Gaussian distribution. More details about this parameter are described in \citet[Section 2.7]{Rubin2015}.}

\added{We use a simple Gaussian mixture model in magnitude for modeling the outlier distribution \citep[c.f.][]{Kunz2007}. Thus, we have no explicit outlier rejection, but as \sne get further from their predicted rest-frame $B$-band magnitudes, they are more and more likely to be described by the outlier distribution. We fix the width of the outlier Gaussian to 0.5 magnitudes, added in quadrature with the measurement uncertainties, and allow the fraction of \sne in this distribution to be a model parameter. A further explanation is presented in \citet[Section 2.3 of ][]{Rubin2015}.}

Using the \snemoseven model with UNITY requires a total of eight standardization coefficients: six for the light-curve-shape eigenvectors, one for the color law, and finally a coefficient describing the effect (if any) of host galaxy stellar mass. These can be combined into a standardized distance modulus equation, following the Tripp convention:
\begin{equation}\label{eqn:unity}
    \mu = m_B - \Big(M_B + \beta A_s + \gamma m + \sum_{i=1}^N \alpha_i c_i\Big)
\end{equation}
where $\mu$, $m_B$, $M_B$ are the distance modulus, apparent and absolute magnitude respectively, the same as \cref{eqn:salt}. For \salt and \snemotwo, $N=1$, but for \snemoseven, $N=6$.
$A_s$ and $\beta$ are the color term and color standardization coefficient respectively. $A_s$ is a spectral variant of the traditional $A_{\rm V}$ extinction. For comparison to \salt, $A_s$ should be approximately $(R_V + 1) c$, meaning that $\beta$ should be $\sim 1$. 
Finally, $\gamma$ is the standardization coefficient applied to the logarithm of the stellar host galaxy stellar mass ($m$).\footnote{When \replaced{standardizing with a mass step}{accounting for host mass using a step function}, as opposed to the linear method presented above, it is common to use $\delta$ as the standardization variable. Host galaxy stellar mass will never be more than a proxy for an astrophysical systematic, and since we are not performing any cosmological measurements, the linear standardization via stellar mass is sufficient even through \replaced{there are more accurate correlations}{more significant correlations may exist} \citep{Childress2014,Rigault2015,Rigault2018,Rose2019}.} The zero point of $m$ is shifted such that the data set's average is zero, partially decorrelating $\gamma$ and the absolute magnitude $M_B$. As this standardization equation uses the same sign for all of the coefficients, these $\alpha$ coefficients have the opposite signs as the one in \cref{eqn:salt}. Due to the small sample sizes, we ran \unity without estimating selection effects or calibration offsets between data sets.

\subsection{\salt Fit as a Reference}\label{sec:salt-ref}

\begin{figure*}
    \centering
    \includegraphics[width=\textwidth]{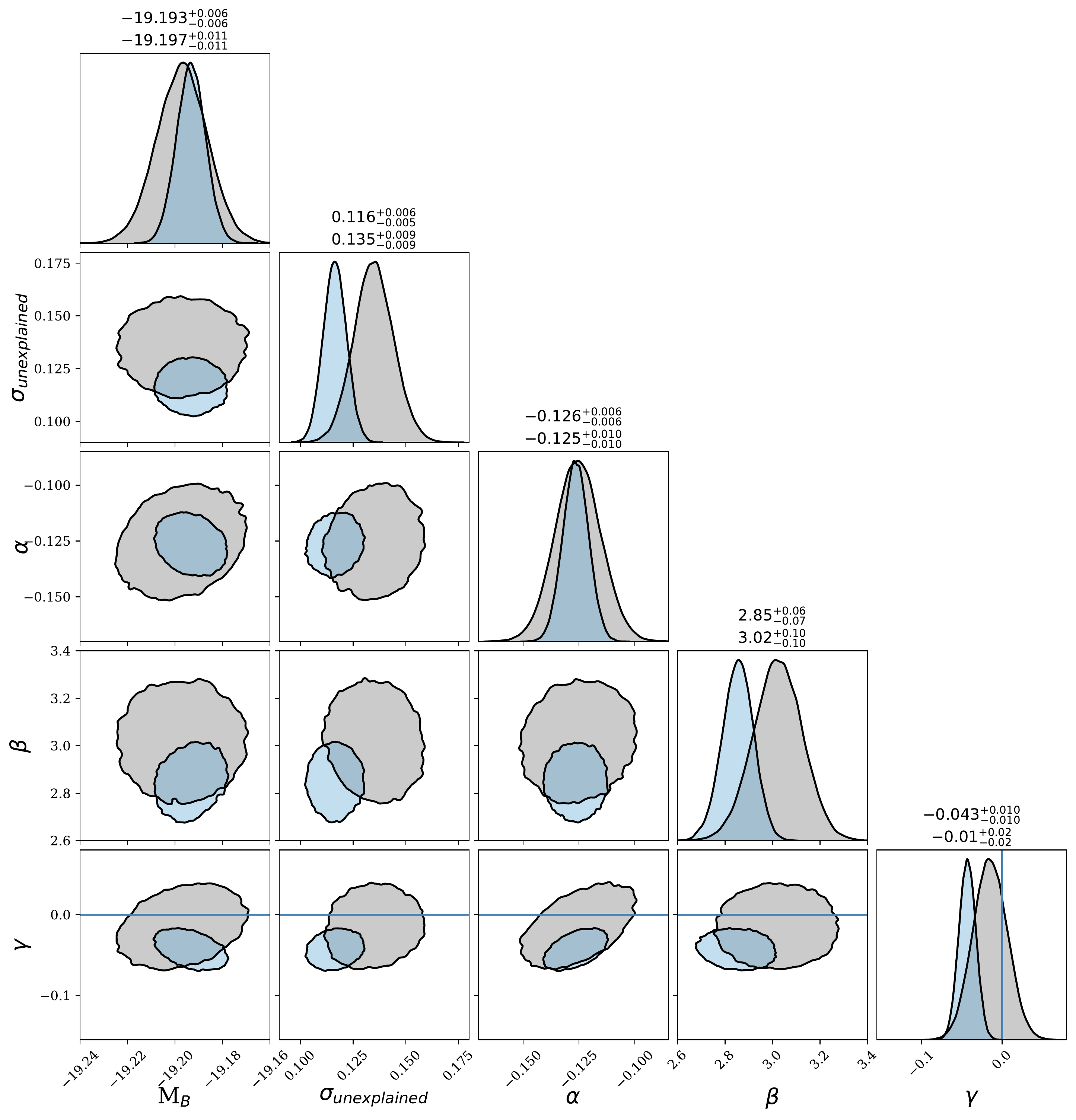}
    \caption{A corner plot of the posterior distribution for the \salt standardization parameters:
    the peak luminosity of \sn (${\rm M}_B$), the unexplained \added{intrinsic} scatter in magnitudes ($\sigma_{\rm unexplained}$), and the \deleted{light-curve} standardization coefficients ($\alpha$, $\beta$, and $\gamma$ from \cref{eqn:unity}).
    Grey contours are for the data set that passed the $\sigma_i \leq 2$ quality cuts with \snemoseven ($N = \nsne$) whereas the blue contours are for the \sn that passed the same cuts when fit with \snemotwo ($N = \nsneSaltSnemoTwo$). Each marginalized distribution's median, along with $1\sigma$ uncertainties, are numerically represented above the corresponding histogram; top and bottom numbers are for the blue and grey distributions, respectively.
    The location of a null host galaxy standardization is shown via the blue line.
    All two-dimensional contours show 2$\sigma$ confidence regions.
    \replaced{These}{This} posterior distribution \replaced{are}{is} consistent with previously published estimates \citep{Sullivan2010,Betoule2014,Scolnic2018}.
    }
    \label{fig:salt_mass}
\end{figure*}

In order to test if \snemo can improve the Hubble-Lema\^itre diagram unexplained dispersion or reduce the correlations with host-galaxy properties, we first need a baseline \replaced{for comparison.}{for our comparison.} As such, we use the \salttwofour version of \salt to fit the \sne that passed \deleted{the} basic quality cuts \deleted{of $\sigma_i \leq \sigmaiNorm$ and $|c_i| < \ciMax$ }for \snemotwo and \snemoseven. The results were then put into \unity to estimate the standardization coefficients \replaced{for}{of} \cref{eqn:unity}. The nominal value for a \salt Tripp-like mass standardization parameter, with the data set that can constrain \snemoseven, was found to be $\gamma = \saltdelta$. 
\added{This is in agreement with $\gamma = 0.042 \pm 0.013$ seen in \citet{Sullivan2010}. Note our inflated uncertainties due to the smaller sample size. When looking at the \sne that can constrain \snemotwo, we measure a nominal value of $\gamma = \saltdeltaSnemoTwo$, nearly identical to that of \citet{Sullivan2010}. This supports our conclusion that our high threshold for light-curve quality does not introduce large biases in the standardization parameters.}
The full fit can be seen in \cref{fig:salt_mass} with the numerical values presented in \cref{tab:snemo2results,tab:snemo7results}. \deleted{We carried out a similar analysis for the larger subset of \sne that passed the \snemotwo quality cuts, finding that the estimated parameters were in agreement with the ones found using the smaller subset.}

\deleted{These results are additionally in agreement with previous measurements from larger data sets, specifically \citet{Sullivan2010}, \citet{Betoule2014}, and \citet{Scolnic2018}. This supports our conclusion that our high threshold for light-curve quality does not introduce large biases in the cosmological parameters. For the host galaxy stellar mass effect, there is no direct literature comparison using a linear standardization and \salttwofour. Our value does agree with the $\gamma = 0.042 \pm 0.013$ seen in \citet{Sullivan2010}, which used SALT2.2.\footnote{\salttwofour improved the size and calibration of the training sample, but there were no significant changes between the two versions.}}

\section{Results \& Discussion} \label{sec:Results}

\subsection{\snemotwo}

\begin{figure*}
    \centering
    \includegraphics[width=\textwidth]{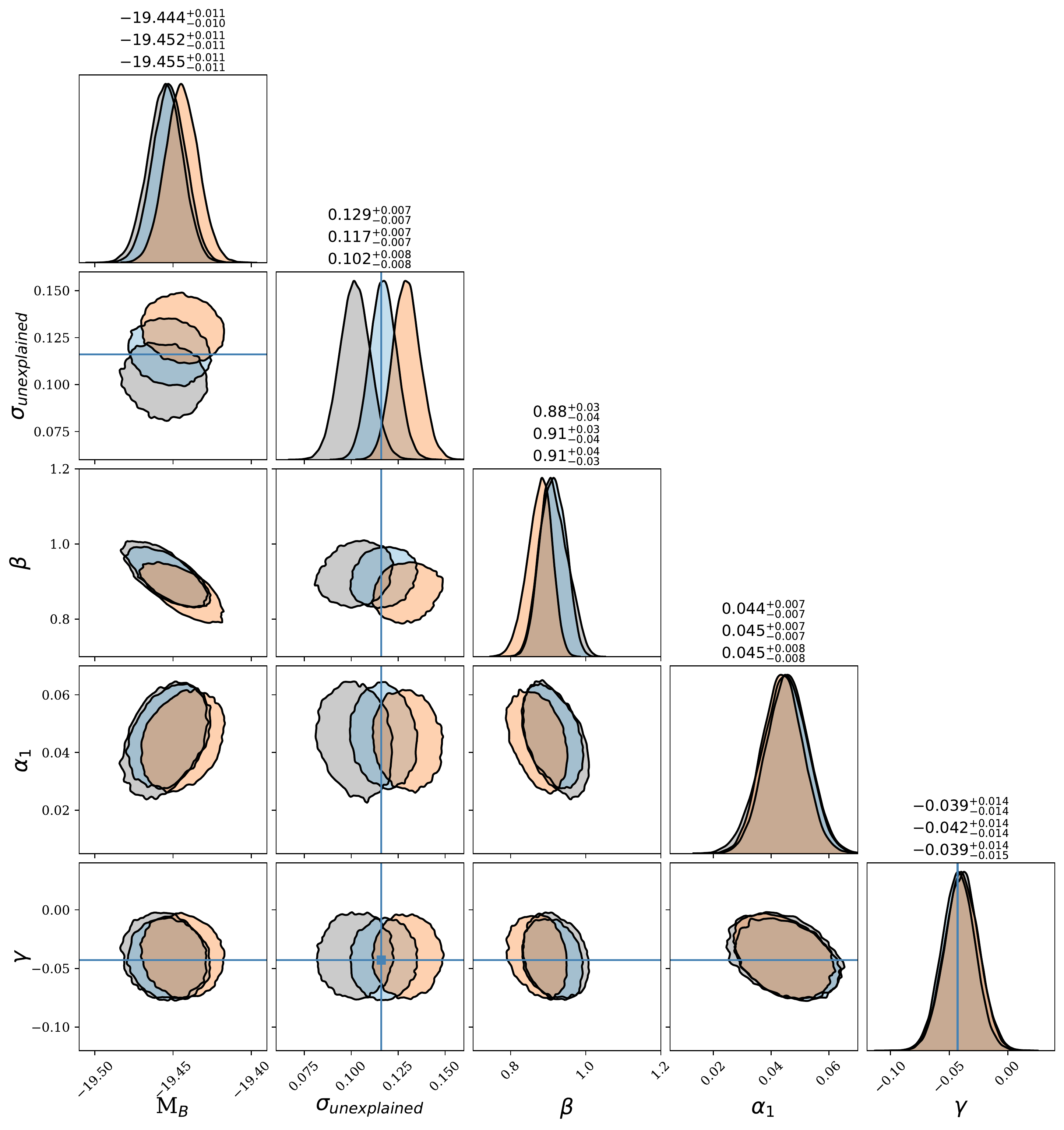}
    \caption{Same as \cref{fig:salt_mass}, but using the \snemotwo light curve fitter. Three different uncertainty models are shown: no uncertainty model (orange), 1\% of peak uncertainty floor (blue), and a 2\% uncertainty (gray).
    Each parameter is well constrained independent of the assumed uncertainty model. As expected, the \added{unexplained} intrinsic scatter ($\sigma_{\rm unexplained}$) depends directly on the uncertainty model. When the uncertainty model decreases to zero (orange contours) the unexplained \added{intrinsic} scatter increases to compensate. The stellar mass dependence is very consistent with what is \replaced{seeing}{seen} when using \salt. The blue lines for the $\sigma_{\rm unexplained}$ and $\gamma$ are the medians of the \salt analysis of the same data set.
    Like \cref{fig:salt_mass}, the median and $1\sigma$ uncertainties for the marginalized distributions are numerically presented above the associated histograms, first for the no uncertainty model (orange), then for the 1\% of peak uncertainty floor (blue), and finally for the 2\% uncertainty floor (gray).
    }
    \label{fig:snemo2}
\end{figure*}

Our first objective was to test the modeling methodology used \replaced{in the}{by} \snemo \deleted{model}. \deleted{\snemotwo was built using the \snemo model assumptions, but only has one light-curve shape parameter and one color term.} With only two model parameters to fit, \snemotwo allows for a direct comparison between \salt and \snemo. \Cref{fig:snemo2} shows the resulting posterior, as inferred by \unity, for the \snemotwo standardization equation. Numerical values are presented in \cref{tab:snemo2results}. 

\snemotwo's standardization parameters are well constrained and independent of the uncertainty model. Each data set analyzed corresponds to one of the three different uncertainty models, uses our default quality cuts of $\sigma_i \leq \sigmaiNorm$ and $|c_i| < \ciMax$, and includes more than 800 \sn. There are no previous measurements for the \snemotwo $\alpha_1$ or $\beta$, but $\sigma_{\rm unexplained}$ and $\gamma$ can be compared to the values from \salt. 
\added{To calculate $\sigma_{\rm unexplained}$ you need to first remove the scatter characterized by the uncertainty model ($\Sigma_{\mathrm{mod}}$, Equation \ref{eq:chisquared}). For even a modest uncertainty model, \snemotwo and \salt have a comparable unexplained intrinsic scatter.}
\deleted{Even without an uncertainty model, $\sigma_{\rm unexplained}$ from \snemotwo is consistent with the value from \salt \replaced{with an}{and its} uncertainty model. We consider this to be an improvement over \salt because an uncertainty model, by definition, is added to explain more of the scatter.
After adding the uncertainty models, giving a more direct comparison to \salt, $\sigma_{\rm unexplained}$ falls to \intrinicScatterSnemoTwoTwoError.}
\replaced{The mass effect}{In addition, the correlation with stellar mass} is not statistically different\deleted{ between \salt and \snemotwo}. Finally, 2--3\% of the \sn were flagged as cosmological outliers.

\replaced{As}{Since} \snemotwo is \replaced{as about as good as}{comparable to} \salt when standardizing \sne, we claim that the modeling details in the \snemo family of models, e.g. wavelength coverage, use of factor analysis, etc, are well-behaved. Following the idea that \sne exhibit more diversity than can be captured by two parameters \citep{Branch2006,Kim2014,Fakhouri2015,Hayden2019,Rubin2019}, we proceed to test the seven parameter \snemoseven model.

\subsection{\snemoseven}

\begin{figure*}
    \centering
    \includegraphics[width=\textwidth]{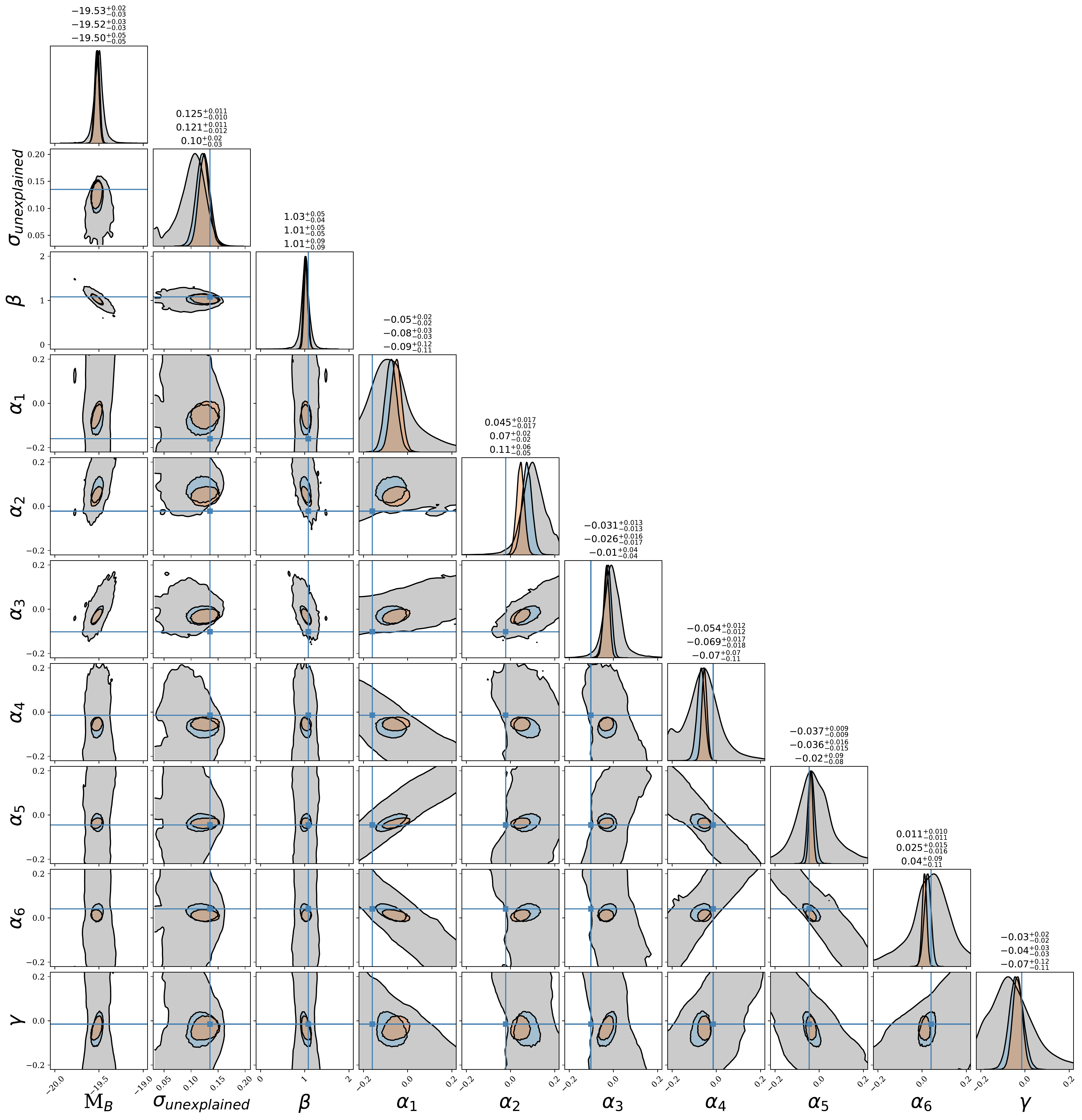}
    \caption{Same as \cref{fig:snemo2}, but using \snemoseven to fit the light curves. Three different uncertainty models are shown: no uncertainty model (orange, top numbers), 1\% of peak uncertainty floor (blue, middle numbers), and a 2\% uncertainty (gray, bottom numbers). \deleted{All data sets use the $\sigma_i \leq 2$  quality cuts.} For several parameters, there is not enough statistical significance to distinguish the standardization parameters ($\alpha_i$, $\gamma$) from zero with any significance, this is especially true for the 2\% uncertainty floor. Nevertheless, the 2\% uncertainty floor does reveal many tight correlations between the parameters (e.g. $\alpha_1$--$\alpha_5$ \& $\alpha_4$--$\alpha_6$) implying that with a reduction of one or two parameters, the others would likely be determinable. 
    The blue lines for $\sigma_{\rm unexplained}$ and $\gamma$ are from the \salt fits of the same data set. Whereas, the other lines, $\alpha_i$ and $\beta$, are the values presented in \citet{Saunders2018}.
    }
    \label{fig:snemo_mass}
\end{figure*}

The results of the analysis of \snemoseven and \unity
are shown in \cref{fig:snemo_mass}. This figure shows the posterior distributions when using the published version of \snemoseven (with no uncertainty model), as well as with the addition of a 1\%, and a 2\% peak luminosity uncertainty model. A 3\% uncertainty model was also tested, but resulted in an exaggeration of the trends already observed \replaced{in}{when} increasing the uncertainty from 1\% to 2\%, and as such is not presented.

There are a few things that stand out from these results. 
First, in \cref{tab:snemo7results}, we see that the outlier percentage typically ranges from \percentOutliersMin to \percentOutliersTypicalMax. 
\unity probabilistically separates these into an outlier population, where they do not affect the inlier population variables: ${\rm M}_{\rm B}$, $\sigma_{\rm unexplained}$, $\alpha_i$, $\beta$, and $\gamma$.
Next, the \added{unexplained} intrinsic scatter starts at \intrinicScatterSevenZeroTwo, slightly smaller than that of \salt, and decreases \added{to \intrinicScatterSevenTwoTwo} if we raise the uncertainty floor to 2\%\deleted{ (\intrinicScatterSevenTwoTwo)}.
Finally, as the size of the uncertainty model increases, the sample size decreases and as expected the uncertainty in the standardization parameters increase.

\begin{deluxetable*}{c||c||ccc}[t]
\tablecolumns{5}
\tablewidth{0pt}
\tablecaption{Parameter estimation results from \unity for \sn that passed \snemotwo quality cuts.\label{tab:snemo2results}} 
\tablehead{
    \colhead{} &
    \colhead{\salt} &
    \multicolumn{3}{c}{\snemotwo}
    \\ 
    \cline{4-4}
    \colhead{\% Error Model}
    & \colhead{} &
    \colhead{0} & \colhead{1} & \colhead{2}
    }
\startdata
Data set size & 867\tablenotemark{\footnotesize a} & 885\tablenotemark{\footnotesize a} & 886 & 881 \\ \hline
M$_B$                       & $-19.193^{+0.006}_{-0.006}$ & $-19.444^{+0.011}_{-0.010}$   & $-19.452^{+0.011}_{-0.011}$     & $-19.455^{+0.011}_{-0.011}$   \\
$\sigma_{\rm unexplained}$  & $0.116^{+0.006}_{-0.005}$   & $0.129^{+0.007}_{-0.007}$ & $0.117^{+0.007}_{-0.007}$  & $0.102^{+0.008}_{-0.008}$ \\
$\beta$                     & $2.85^{+0.06}_{-0.07}$      & $0.88^{+0.03}_{-0.04}$    & $0.91^{+0.03}_{-0.04}$   & $0.91^{+0.04}_{-0.03}$ \\
$\alpha_1$                  & $-0.126^{+0.006}_{-0.006}$  & $0.044^{+0.007}_{-0.007}$ & $0.045^{+0.007}_{-0.007}$   & $0.045^{+0.008}_{-0.008}$ \\
$\gamma$                    & $-0.043^{+0.010}_{-0.010}$  & $-0.039^{+0.014}_{-0.014}$ & $-0.042^{+0.014}_{-0.014}$ & $-0.039^{+0.014}_{-0.015}$ \\
No. of outliers & 15 (1.7\%) & 21 (2.4\%) & 20 (2.3\%) & 26 (3.0\%) \\
\enddata
\tablecomments{The \sn used in the \salt analysis are the ones that passed the $\sigma_i \leq 2$ for \snemotwo with no error model and were successfully fit with \salt. The ``No. of outliers'' is reported both as an absolute number and a percentage of the data set.}
\tablenotetext{a}{\added{For the same initial data set, \snemotwo has 6 additional \sne rejected as cosmological outliers, where as 18 additional \sne that were rejected at the \salt light-curve fitting stage.}}
\end{deluxetable*}

\begin{deluxetable*}{c||ccc||c||ccc}[t]
\tablecolumns{8}
\tabletypesize{\footnotesize}  
\tablewidth{0pt}
\tablecaption{Parameter estimation results from \unity for \sn that passed \snemoseven quality cuts\added{, $\sigma_i \leq 2$}.\label{tab:snemo7results} \explain{All results for $\sigma_i \leq 1$ data sets are now all presented together in the appendix.}} 
\tablehead{
    \colhead{} &
    \multicolumn{3}{c}{\salt} &
    \colhead{\citet{Saunders2018}} & 
    \multicolumn{3}{c}{\snemoseven}
    \\ 
    \cline{3-3}
    \cline{7-7}
    \colhead{\% Error Model}
    & \colhead{} & \colhead{} & \colhead{} & \colhead{} &
    \colhead{0}  & \colhead{1} & \colhead{2}
    }
\startdata
Data set size & 240 & 194 & 126 & 133 & 240 & 194 & 126 \\ \hline
M$_B$  & $-19.197^{+0.011}_{-0.011}$                      & $-19.196^{+0.012}_{-0.012}$                      & $-19.188^{+0.017}_{-0.017}$                      & \nodata           & $-19.53^{+0.02}_{-0.03}$ & $-19.52^{+0.03}_{-0.03}$  & $-19.50^{+0.05}_{-0.05}$  \\
$\sigma_{\rm unexplained}$  & $0.135^{+0.009}_{-0.009}$ & $0.131^{+0.011}_{-0.011}$ & $0.141^{+0.015}_{-0.015}$ & \nodata           & $0.125^{+0.011}_{-0.010}$ & $0.121^{+0.011}_{-0.012}$ & $0.10^{+0.02}_{-0.03}$\\
$\beta$  & $3.02^{+0.10}_{-0.10}$                & $2.84^{+0.10}_{-0.10}$                & $2.96^{+0.14}_{-0.14}$                & $1.08 \pm 0.04$    & $1.03^{+0.05}_{-0.04}$ & $1.01^{+0.05}_{-0.05}$ & $1.01^{+0.09}_{-0.09}$ \\
$\alpha_1$  & $-0.125^{+0.010}_{-0.010}$             & $-0.129^{+0.012}_{-0.012}$             & $-0.122^{+0.017}_{-0.017}$             & $0.16 \pm 0.03$    & $-0.05^{+0.02}_{-0.02}$ & $-0.08^{+0.03}_{-0.03}$ & $-0.09^{+0.12}_{-0.11}$\\
$\alpha_2$  & \nodata                                                                        & \nodata                                                                        & \nodata                                                                        & $0.02 \pm 0.03$    & $0.045^{+0.017}_{-0.017}$ & $0.07^{+0.02}_{-0.02}$ & $0.11^{+0.06}_{-0.05}$\\
$\alpha_3$  & \nodata                                                                        & \nodata                                                                        & \nodata                                                                        & $0.103 \pm 0.017$  & $-0.031^{+0.013}_{-0.013}$ & $-0.026^{+0.016}_{-0.017}$ & $-0.01^{+0.04}_{-0.04}$\\
$\alpha_4$  & \nodata                                                                        & \nodata                                                                        & \nodata                                                                        & $0.01 \pm 0.02$    & $-0.054^{+0.012}_{-0.012}$ & $-0.069^{+0.017}_{-0.018}$ & $-0.07^{+0.07}_{-0.11}$\\
$\alpha_5$  & \nodata                                                                        & \nodata                                                                        & \nodata                                                                        & $0.045 \pm 0.009$  & $-0.037^{+0.009}_{-0.009}$ & $-0.036^{+0.016}_{-0.015}$ & $-0.02^{+0.09}_{-0.08}$\\
$\alpha_6$  & \nodata                                                                        & \nodata                                                                        & \nodata                                                                        & $-0.041 \pm 0.017$ & $0.011^{+0.010}_{-0.011}$ & $0.025^{+0.015}_{-0.016}$ & $0.04^{+0.09}_{-0.11}$\\
$\gamma$  & $-0.01^{+0.02}_{-0.02}$             & $-0.03^{+0.03}_{-0.03}$             & $-0.04^{+0.05}_{-0.05}$             & \nodata           & $-0.03^{+0.02}_{-0.02}$ & $-0.04^{+0.03}_{-0.03}$ & $-0.07^{+0.12}_{-0.11}$\\
No. of outliers & 4 (1.7\%) & 4 (2.1\%) & 4 (3.2\%) & \nodata & 7 (2.9\%) & 3 (1.5\%) & 4 (3.2\%)\\
\enddata
\tablecomments{The \sn used in the \salt analysis are the ones that passed \replaced{the $\sigma_i \leq 2$ for \snemoseven with no error model.}{each of the \snemoseven analyses, respectively.} The ``No. of outliers'' is reported both as an absolute number and a percent of the data set.}
\end{deluxetable*}

\subsubsection{\qoneshort}


\Cref{tab:snemo7results} shows the estimated standardization coefficients after applying various data quality cuts. 
Taking the data with no uncertainty model added, we determine that they differ at $\sim 2 \sigma$ with the original estimates produced when using the SNfactory data set \citep[and forthcoming erratum]{Saunders2018}. When evaluating seven parameters, it is expected to see some variability. In this analysis, $\alpha_3$ differs at $>3\sigma$ from the SNfactory numbers. Having only one parameter reach this level of disagreement is expected in about 2\% of analysis, or $\gtrsim 2 \sigma$.
Including a non-zero mass standardization \deleted{(as we have done, but \citealt{Saunders2018} did not)} does slightly shift the central values of the other standardization coefficients, but not by the scale of the variation described above. The correlations between $\gamma$ and each $\alpha$, seen in \cref{fig:snemo_mass}, are not large enough to cause a drastic shift in any of the standardization coefficients. 
Furthermore, these values can shift by over $1\sigma$ (e.g. $\alpha_1$) with the addition of a 1\% uncertainty model. 
\replaced{To conclude, o}{O}ur results show that the standardization coefficients for \snemoseven show only mild variation between data sets.

\subsubsection{\qtwo}

Using \snemoseven with no model uncertainties, most coefficients can be distinguished from zero at $>2\sigma$, with $\alpha_4$ distinguishable from 0 at greater than $4\sigma$. A 1\% uncertainty model has similar results.
However with a 2\% uncertainty model, \unity is unable to distinguish the \snemoseven standardization components from zero (except for $\alpha_2$). This is likely due to a combination of data set size and the quality of the light curves themselves. Assuming \snemoseven has an uncertainty model below $\sim$ 2\%, each light-curve parameter will have a non-zero standardization coefficient.

\subsubsection{\qthree}

\replaced{With a 2\% uncertainty model, it is impossible to differentiate the standardization coefficients from zero. However, these results do}{The 2\% uncertainty model does} reveal strong correlations between the parameters.
\replaced{Do to t}{T}hese strong correlations \added{suggest that} the constrainability of the standardization parameters would dramatically improve if one or two of these parameters were fixed or known.
A lower dimensional model (like \snemosix or \snemofive) would have an effect similar to ``fixing'' one or two of these parameters to zero. However, since a five parameter EMFA model is not simply the first five parameters of a seven parameter EMFA model, \deleted{this would only be a partial solution. In addition, this property of EMFA models means that making} \snemofive \replaced{requires}{would require a full} retraining rather than \added{a simple} truncati\replaced{ng}{on of} \snemoseven.

When looking at spectral time series data, \snemoseven appears to be a viable photometric light-curve fitter, but these strong correlations imply that not all of the eigenvectors are constrainable with today's light curves. 
Since a much higher percentage of the higher cadence CSP \sn passed quality cuts, we know that the quality of the observed light curves \added{(as measured by wavelength coverage, signal-to-noise, temporal sampling, etc.)} plays a role in the ability of \snemoseven to be used with photometric data.
\replaced{Alternatively}{Additionally}, the eigenvectors that could be obtained from light curves are not necessarily the same, nor in the same order, as those obtained from spectral time series (like the \snemo eigenvectors).
Similar to the work of \citet{Kim2013}, the \snemo eigenvectors manifested in light curves should be investigated and perhaps a new model generated that prioritizes the information available in the light curves.

\subsubsection{\qfourfive}

The final two questions deal with the uniformity of the standardization.
Using the \added{\snemoseven} light curves \deleted{that pass the $\sigma_i\leq 2$ cuts for \snemoseven} with no additional error model, the unexplained intrinsic scatter ($\sigma_{\rm unexplained}$) moderately decreased from \intrinicScatterSaltZeroTwo with \salt to \intrinicScatterSevenZeroTwo, \added{for the same \sne}. We found that with a 2\% uncertainty model, the unexplained intrinsic scatter decreased from \intrinicScatterSaltTwoTwo for \salt to \intrinicScatterSevenTwoTwo.
Because this is a more direct comparison to the $\sigma_{\rm unexplained}$ of \salt, as both methods use some uncertainty model, we conclude that \snemoseven is capable of\deleted{, though not statistically significant with this data set,} decreasing the unexplained intrinsic scatter on the Hubble-Lema\^itre diagram. With regard to the reduction of unexplained variation or systematic limits of standardization, \snemoseven shows only slightly significant deviations from \snemotwo or \salt.

We also found that there was no apparent decrease in the host galaxy stellar mass dependence. For the \snemoseven data set, \salt only sees a non-zero mass dependence at $0.5\sigma$, whereas, with no uncertainty model \deleted{and a $\sigma_i \leq 2$ quality cut,} the mass dependence of \snemoseven was measured to have a $1.5\sigma$ non-zero statistical significance.
\replaced{But this shift is not statistically significant.}{As discussed in \cref{sec:salt-ref}, this is not a removal of a host galaxy mass correlation, but more likely an inflation of its uncertainty due to the small sample size.}

\section{Impacts on Systematics Dominated Cosmology}\label{sec:impacts}

\added{Current cosmological analyses typically use \salt as their nominal model for standardizing \sn magnitudes \citep[e.g.][]{Scolnic2018,DESCollaboration2019}. 
The next generation of \sn cosmological surveys will be limited by systematic uncertainties.
\snemoseven's decreased scatter (RMS) in \sn absolute magnitude, as compared to \salt \citep{Saunders2018}, is expected to reduce this systematic uncertainty floor. The above analysis tests if \snemo can be used on the same data currently used with \salt.}

\added{As seen in \cref{fig:snemo2}, \snemotwo can be used as a drop in replacement for \salt. For a full cosmological analysis, \snemotwo would need to be merged into current cosmological tools \citep[e.g.][]{Kunz2007,Kessler2009a,Kessler2017}. In addition, we expect \snemotwo to get the minor revisions and improvements \salt has received over the last 13 years. For example, \snemotwo will benefit from the linking of SNFacotry to the CALSPEC system \citep[Rubin et al. in prep.]{Bohlin2014}.}


\begin{figure}
    \centering
    \includegraphics[width=0.95\columnwidth]{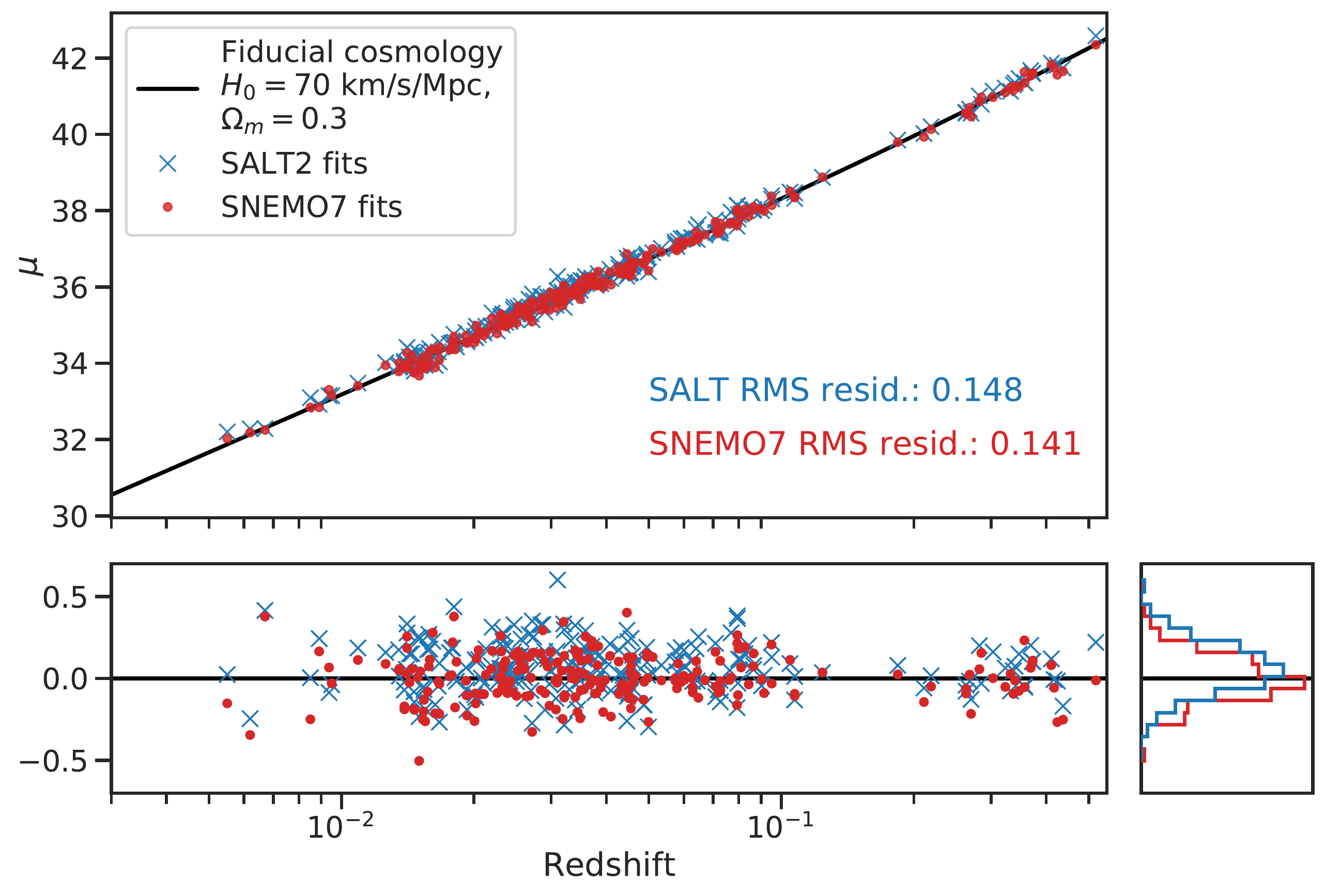}
    \caption{\added{The Hubble-Lema\^itre diagram using \salt (blue Xs) and \snemoseven (red points).
    As seen in \citet{Saunders2018}, the scatter for the same \sne is smaller when using \snemoseven.
    This figure assumes a fiducial cosmology and does not correct for any Malmquist biases.}
    }
    \label{fig:HL-diagram}
\end{figure}

\begin{figure}
    \centering
    \includegraphics[width=0.95\columnwidth]{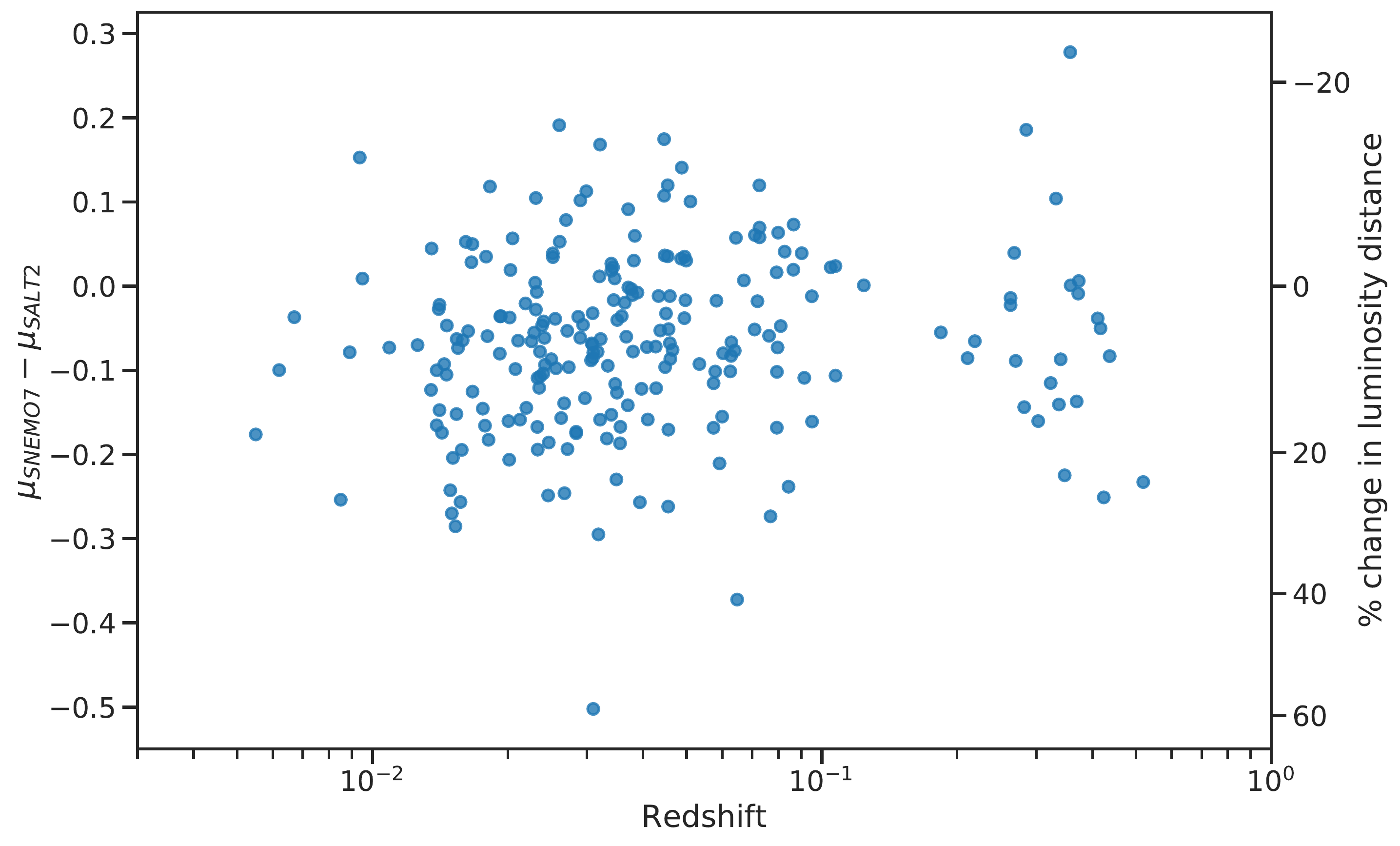}
    \caption{\added{
    The difference in distance between \salt and \snemoseven vs redshift.
    There is no significant redshift dependence.
    However, \snemoseven does increase the average \sn modeled distance (negative average $\Delta \mu$). This would be a change in M$_B$ but would not effect the estimation of dark energy. }
    }
    \label{fig:deltamu}
\end{figure}

\added{Although a full cosmological treatment is not possible, if a fiducial cosmology is assumed, comparisons between \salt{} and \snemoseven can be made. \Cref{fig:HL-diagram} shows a Hubble-Lema\^itre diagram using both the \salt and \snemoseven models, assuming the same fiducial cosmology as \unity. The data shown are from the nominal sample, with $|c_i| < \ciMax$, $\sigma_i \leq \sigmaiNorm$, and the \snemoseven model being fit with no added error model. We additionally cut all objects identified as outliers by \unity, leaving 229 \sne.}

\added{The resulting Hubble-Lema\^itre diagram confirms the result from \citet{Saunders2018} that \snemoseven reduces the scatter around the assumed cosmology from an RMS of \saltrms to \snemorms. 
These RMS values are from the mean of each residual distribution, not the zero fiducial cosmology itself.
Though the extra degrees of freedom in \snemoseven may be unexpectedly self-serving, this is unlikely because the data set has both \snemoseven and \salt outliers rejected.
In addition, this small reduction in overall scatter takes on a larger significance since these numbers are for the same \sne. 
}

\added{The difference in distance between \salt and \snemoseven verse redshift can be seen in \cref{fig:deltamu}, where no redshift dependence is visible.
The negative average $\Delta \mu$ indicates a change to the model M$_B$ value, as seen in Table \ref{tab:snemo7results}. A lack of redshift decadence indicates that there would be no effect on any dark energy parameters.
}

\added{A full cosmological analysis with \snemoseven is not yet possible, but the reduced RMS in the Hubble-Lema\^itre diagram shows promise that these new models can explain more \sn variation than \salt. Explaining more variation is important because any unaccounted variance may produce a systematic offset between \sn at different redshifts. This possibility is a major systematic uncertainty for future cosmological measurements.}

\added{Continued work is required to improve \snemo and other new light curve fitters, particularly since only $\sim 25\%$ of available photometrically observed \sne can use \snemoseven. An error model is coming to \snemo, as is further testing of the necessary data quality. These improvements are warranted because \citet{Saunders2018} and this work have shown that improvements on \salt are possible. However, there is currently no model that can act as a drop in replacement while giving us these improvements.}

\section{Conclusion}

\snemo is a family of \sn models trained on the spectrophotometric time series data of SNfactory. One of the many potential uses of these models is to standardise \sne for cosmological measurements. In testing that use case with current data sets, we are able to consistently determine the standardisation parameters for \snemoseven, but tight correlations between the parameters implies that with a reduction of \added{the model complexity by} one or two \added{components}, the other parameters should be much easier to \replaced{determine}{constrain}. This means that \snemofive or \snemosix would be good candidates for a new light-curve fitter with a goal of cosmological standardisation of current data sets.


\deleted{As currently published \salt, \snemotwo, and \snemoseven all have similar $\sigma_{\rm unexplained}$.} \added{To properly calculate $\sigma_{\rm unexplained}$, you need to first remove the scatter characterized by the uncertainty model.} Once we add a \replaced{basic}{modest} uncertainty model --- \replaced{like already}{similar to the one} present in \salt{} --- \snemotwo and \snemoseven have a \added{slight} reduction in $\sigma_{\rm unexplained}$ indicating that \snemo explains more of the natural variation of \sn. On the other hand, there is no statistically significant reduction in a stellar mass dependence, implying that adding more linearly standardized light-curve parameters, with \snemoseven, would be susceptible to a similar systematic uncertainty floor as \salt and that we may be approaching the limits of Tripp-standardization. This perhaps motivates consideration of non-linear relationships. \explain{This was true in earlier runs, but is not obvious in the final runs. It was removed from Section \ref{sec:Results} but was accidentally left into the conclusion prior to submission.}\deleted{Looking at the specific correlation between $\sigma_{\rm unexplained}$ and $\gamma$, \salt shows a slight correlation compared to \snemoseven's no correlation. This implies that at least part of the information parameterized by the host stellar mass is contained in the \snemoseven eigenvectors.}

\snemoseven describes more of the intrinsic variation of \sn\replaced{, and these differences}{ as seen in the reduction of the RMS in Figure \ref{fig:HL-diagram}. These unaccounted for variations} have been shown to be responsible for a significant fraction of the unexplained intrinsic dispersion seen in \salt analyses \citep{Fakhouri2015}. Therefore, there is a danger that if one leaves these differences unaccounted for, \sn sets at different redshifts could systematically favor one side or the other of this unexplained intrinsic dispersion, thus introducing a systematic in any cosmological measurement. If we cannot constrain models that explain more of this intrinsic dispersion, we risk being unable to reach the level of precision planned for future cosmological surveys. 

The family of \snemo models are not intrinsically unconstrainable, as \snemotwo can easily be constrained with present data. On the other hand, 
only the highest quality among currently available light curves could be fit by \snemoseven,
and the resulting data set is dominated by \sn observed by CSP. Therefore, it is likely that upcoming large surveys, such as LSST and WFIRST, will want to specify CSP-like signal-to-noise, time sampling, and rest-frame wavelength coverage for a reasonable fraction of their supernova photometry. Such light curves could be fit by \snemoseven and gain the benefit of better constraints on the \sn differences. It is also possible that one or more spectra might be needed. Further work is essential in order to properly understand the data requirements needed for high quality and cosmologically useful fits of light curves with \snemoseven.
Part of this work is already in preparation but additional \snemo models with fewer light-curve shape parameters should also be investigated. This would include an investigation into possible information loss or reordering of eigenvectors by going from the spectrophotmetric time series data to light-curve data.

\replaced{This paper is a first look at \snemoseven before it is used in cosmological analyses. If \snemoseven could be used as a drop-in replacement for \salt with the current data and standardization methods, there would be no further work needed.}{We have presented a first look at \snemoseven's ability to be a replacement for \salt in cosmological analyses.}
\added{We have concluded that further analyses are required to determine what CSP-like qualities are needed to use the additional information in \snemoseven. Also, \snemo models with fewer parameters should be developed and tested in order to use lower quality data sets,}
\deleted{However, with current data sets and analysis methods, \snemoseven does not improve cosmology measurements, but} since neither the philosophy behind \snemo nor photometry-only data sets are roadblocks\replaced{, specific survey optimization analyses are required in order to use the additional information seen in the extra parameters of \snemoseven.}{ to its future use.}

\acknowledgements
The authors thank Ravi Gupta and David Jones for the stellar masses associated with Foundation and CSP \sn hosts.
In addition, we thank 
Greg Aldering, 
David Rabinowitz, 
and the entire WFIRST Supernova Science Investigation Team (PI: Perlmutter, S.), for comments on early drafts.
\added{We would like to also thank the anonymous referee for their  comments.}
BR, SD, DR, RH SD, AF, LG, SP, and MS acknowledge support from NASA through grant NNG16PJ311I.
CS acknowledges support from the Labex ILP (reference ANR-10-LABX-63) part of the Idex SUPER, and received financial state aid managed by the Agence Nationale de la Recherche, as part of the programme Investissements d'avenir under the reference ANR-11-IDEX-0004-02.
LG was funded by the European Union's Horizon 2020 research and innovation programme under the Marie Sk\l{}odowska-Curie grant agreement No. 839090.

\software{astropy \citep{Astropy}, 
click, 
corner.py \citep{Foreman-Mackey2016}, 
emcee \citep{Foreman-Mackey2012}, 
kde\_corner,
Matplotlib \citep{matplotlib}, 
Numpy \citep{numpy}, 
Pandas \citep{pandas}, 
pystan (\doi{10.5281/zenodo.598257}),
python, 
SciPy \citep{scipy}, 
Seaborn (\doi{10.5281/zenodo.883859}),
sncosmo (\doi{10.5281/zenodo.592747}), 
Stan \citep{Carpenter2017}
}

\appendix
\section{Effect of Signal to Noise Cuts}\label{sec:SNRCut}

\begin{figure*}
    \centering
    \includegraphics[width=\textwidth]{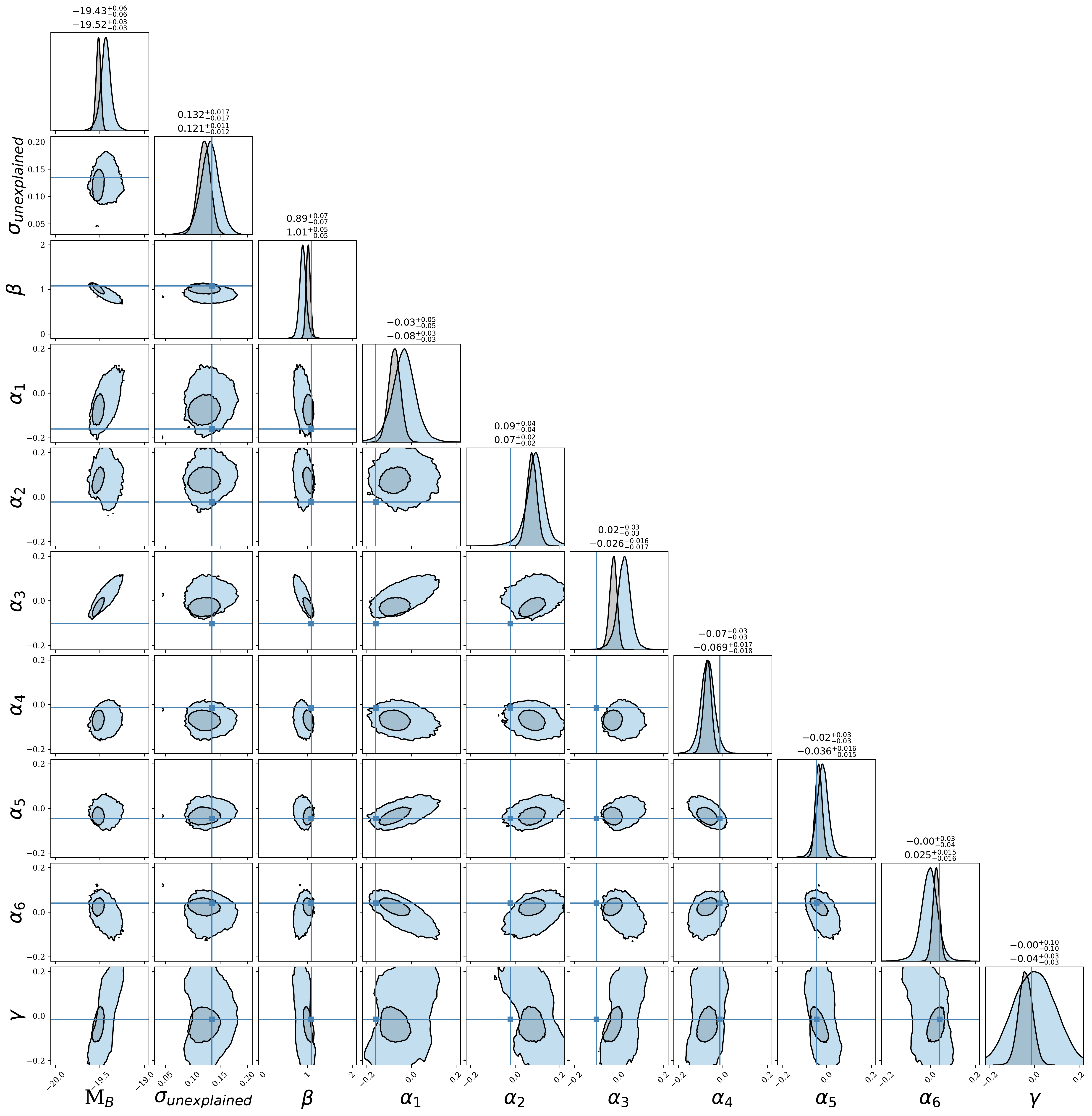}
    \caption{The same as \cref{fig:snemo_mass} but showing the effect of changing $\sigma_i$. These data sets use a 1\% uncertainty model and use quality cuts of $\sigma_i \leq 1$ (blue) and $\sigma_i \leq 2$ (gray, blue in \cref{fig:snemo_mass}). Changing the signal to noise cut has \deleted{is} no significant ($< 2 \sigma$) effect on the \unity parameter estimation. This increase in uncertainty is expected for the decrease in sample size.
    }
    \label{fig:snemo_mass_}
\end{figure*}

The $\sigma_i \leq \sigmaiNorm$ cut is a subjective choice and therefore could have a noticeable effect on the results presented. As such, we reran \unity on data sets with cuts applied at $\sigma_i \leq 1$. 
These results\deleted{, along with all the others,} are listed in \replaced{Table \ref{tab:snemo7results}}{Table \ref{tab:snemo7sigma1results}}. 
While using the 1\% uncertainty model, \cref{fig:snemo_mass_} shows the \replaced{results of these two quality cuts}{effects of changing the value of the $\sigma_i$ cut}. 

As expected, the effect of changing from a quality cut of $\sigma_i \leq \sigmaiNorm$ are slight and statistically insignificant. The uncertainty on the parameters are inflated when moving from $\sigma_i \leq \sigmaiNorm$ to $\sigma_i \leq 1$ but this is largely due to the decrease in sample size, from \nsneOnePercent \sn to \nsneOnePercentOneSigma respectively, rather than the actual value of $\sigma_i$. Ultimately, the ability to standardize \snemoseven light-curve fits shows no significant dependence on reasonable light-curve fit quality cuts.

\added{The 2\% error model in Table \ref{tab:snemo7sigma1results} shows very large uncertainties, particularly for $\gamma$. This is not an issue with \unity, but rather a result of attempting to constrain eight standardization parameters via 31 inlier \sn (39 total minus 8 outliers). In addition, the distribution of \sne in this data set are biased toward more massive hosts, forcing an even larger uncertainty on $\gamma$. These numbers are presented for completeness only.}

\begin{deluxetable}{c|ccc}[t]
\tablecolumns{4}
\tabletypesize{\footnotesize}  
\tablewidth{0pt}
\tablecaption{\added{Parameter estimation results from \unity for \sn that passed \snemoseven quality cuts, $\sigma_i < 1$.\label{tab:snemo7sigma1results}}} 
\tablehead{
    \colhead{} &
    \multicolumn{3}{c}{\snemoseven}
    \\ 
    \cline{3-3}
    \colhead{\% Error Model}
    &
    \colhead{0}  & \colhead{1} & \colhead{2}
    }
\startdata
Data set size & 157 & 90 & 39 \\ \hline
M$_B$  & $-19.52^{+0.04}_{-0.05}$ & $-19.43^{+0.06}_{-0.06}$  & $-19.4^{+0.2}_{-0.3}$  \\
$\sigma_{\rm unexplained}$  & $0.135^{+0.015}_{-0.014}$ & $0.132^{+0.017}_{-0.017}$ & $0.06^{+0.04}_{-0.03}$\\
$\beta$  & $1.05^{+0.07}_{-0.06}$ & $0.89^{+0.07}_{-0.07}$ & $0.71^{+0.17}_{-0.17}$ \\
$\alpha_1$  & $-0.05^{+0.03}_{-0.04}$ & $-0.03^{+0.05}_{-0.05}$ & $0.18^{+0.07}_{-0.07}$\\
$\alpha_2$  & $0.04^{+0.02}_{-0.03}$ & $0.09^{+0.04}_{-0.04}$ & $0.18^{+0.05}_{-0.05}$\\
$\alpha_3$  & $-0.03^{+0.02}_{-0.02}$ & $0.02^{+0.03}_{-0.03}$ & $0.08^{+0.05}_{-0.05}$\\
$\alpha_4$  & $-0.055^{+0.017}_{-0.017}$ & $-0.07^{+0.03}_{-0.03}$ & $-0.17^{+0.05}_{-0.06}$\\
$\alpha_5$  & $-0.040^{+0.014}_{-0.013}$ & $-0.02^{+0.03}_{-0.03}$ & $0.07^{+0.04}_{-0.04}$\\
$\alpha_6$  & $0.014^{+0.016}_{-0.017}$ & $-0.004^{+0.03}_{-0.04}$ & $-0.10^{+0.06}_{-0.06}$\\
$\gamma$  & $-0.06^{+0.05}_{-0.05}$ & $-0.001^{+0.10}_{-0.10}$ & $-0.02^{+0.9}_{-1.1}$\\
No. of outliers & 6 (3.8\%) & 1 (1.1\%) & 8 (21\%) \\
\enddata
\tablecomments{The ``No. of outliers'' is reported both as an absolute number and a percent of the data set.}
\end{deluxetable}

\bibliographystyle{apj}
\bibliography{library}

\listofchanges

\end{document}